\documentclass[lettersize,journal]{IEEEtran}

\usepackage{cite}
\usepackage{amsmath,amssymb,amsfonts}
\usepackage{algorithmic}
\usepackage{array}
\usepackage{textcomp}
\usepackage{stfloats}
\usepackage{url}
\usepackage{verbatim}
\usepackage{graphicx}
\usepackage{xcolor}
\usepackage[inline]{enumitem}
\usepackage{cleveref}
\usepackage{multirow}
\usepackage[T1]{fontenc}
\usepackage{subcaption}
\usepackage{adjustbox}
\usepackage[detect-none]{siunitx}

\def\BibTeX{{\rm B\kern-.05em{\sc i\kern-.025em b}\kern-.08em
    T\kern-.1667em\lower.7ex\hbox{E}\kern-.125emX}}    
\usepackage{balance}
\begin{document}
\title{Aerodynamic Performance and Impact Analysis of a MEMS-Based Non-Invasive Monitoring System for Wind Turbine Blades}
\author{
Nicolas Schärer, 
Denis Mikhaylov, 
Cédric Sievi, 
Badoui Hanna, 
Caroline Braud,
Julien Deparday, 
Sarah Barber, 
Tommaso Polonelli,
Michele Magno 

\thanks{Nicolas Schärer, Denis Mikhaylov, Cédric Sievi, Tommaso Polonelli and Michele Magno are with PBL, ETH Zürich, Zürich, Switzerland}
\thanks{Julien Deparday and Sarah Barber are with IET, OST, Rapperswil, Switzerland}
\thanks{Badoui Hanna and Caroline Braud are with LHEAA, CSTB, Nantes, France}
\thanks{The experiments were performed within the French-Swiss project MISTERY funded by the French National Research Agency (ANR PRCI grant no. 266157) and the Swiss National Science Foundation (grant no. 200021L 21271).}
}


\maketitle

\begin{abstract}
Wind power generation plays a crucial role in transitioning away from fossil fuel-dependent energy sources, contributing significantly to the mitigation of climate change. Monitoring and evaluating the aerodynamics of large wind turbine rotors is crucial to enable more wind energy deployment. This is necessary to achieve the European climate goal of a reduction in net greenhouse gas emissions by at least 55\% by 2030,  compared to 1990 levels. This paper presents a comparison between two measurement systems for evaluating the aerodynamic performance of wind turbine rotor blades on a full-scale wind tunnel test. One system uses an array of ten commercial compact ultra-low power micro-electromechanical systems (MEMS) pressure sensors placed on the blade surface, while the other employs high-accuracy lab-based pressure scanners embedded in the airfoil. The tests are conducted at a Reynolds number of \num[detect-weight=true, detect-family=true, mode=text]{3.5d6}, which represents typical operating conditions for wind turbines. MEMS sensors are of particular interest, as they can enable real-time monitoring which would be impossible with the ground truth system. This work provides an accurate quantification of the impact of the MEMS system on the blade aerodynamics and its measurement accuracy. Our results indicate that MEMS sensors, with a total sensing power below \qty[detect-weight=true, detect-family=true, mode=text]{1.6}{\milli\watt}, can measure key aerodynamic parameters like Angle of Attack (AoA) and flow separation with a precision of \ang[detect-weight=true, detect-family=true, mode=text]{1}. Although there are minor differences in measurements due to sensor encapsulation, the MEMS system does not significantly compromise blade aerodynamics, with a maximum shift in the angle of attack for flow separation of only \ang{1}.
 These findings indicate that surface and low-power MEMS sensor systems are a promising approach for efficient and sustainable wind turbine monitoring using self-sustaining Internet of Things devices and wireless sensor networks.
\end{abstract}


\section{Introduction}
\IEEEPARstart{T}{oday}, wind energy is considered one of the most important resources to significantly reduce CO2 emissions and help mitigate global warming~\cite{kanal2020assessment,rezamand2020critical}. Based on the recently agreed EU target, renewable energy must make up at least 42\% of the electrical grid by 2030~\cite{albulescu2020co}. Thus, a massive increase in wind turbine capacity and efficiency is required, with wind power generation capacity expected to grow from \qty{204}{\giga\watt} in 2022 to more than \qty{500}{\giga\watt} in 2030~\cite{artale2019monitoring}. To achieve this, optimizing the design and operation efficiency of wind turbines while reducing negative environmental impacts and maintenance costs is essential~\cite{artale2019monitoring}. A far deeper understanding of the blade aerodynamics~\cite{fahim2022machine} is needed, specifically for off-shore wind farms, where installation and periodic maintenance costs require a high volume of generated electricity to be commercially viable. 
Wind turbines are exposed to gusts and continual strong wind speed changes over time, which can cause load fluctuations, reduced lifespan, and less effective operational control~\cite{wang2022influence}. 
A better understanding of these unsteady and potentially turbulent flows over airfoils, such as wind turbine blades, would enhance dynamic force estimation and allow more efficient and adaptive airfoil control models~\cite{de2021controlling}. For this purpose, experimental data are necessary, including parameters like Angle of Attack (AoA) and flow separation~\cite{huang2023layered}. The latter is characterized by a separation point at which the airflow no longer follows the shape of the airfoil~\cite{wang2024passive}. As the AoA increases, the separation point moves from the trailing edge towards the leading edge and reduces the aerodynamic performance~\cite{braud2023study}. However, there is currently a lack of available experimental data for such airfoils with controlled unsteady inflow to measure flow structures and surface pressure responses~\cite{hanna2024reynolds} at high Reynolds numbers, i.e., \( Re > 10^6 \). While pressure scanners enable such data to be collected in a wind tunnel, complex calibration and the invasive nature of the pressure taps and fluid pipes inherently limit field deployment on operating turbine blades. Therefore, such pressure scanner systems are not suitable for large-scale installation on operational wind turbines, let alone retrofitting~\cite{barber2022development,polonelli2022aerosense,fischer2021windnode}. 

Wireless Sensor Networks (WSN) hold promise for monitoring wind turbine structures, driven by the global interest in the Internet of Things (IoT)~\cite{vazquez2022energy}. Many wireless systems use low-power, compact Micro-ElectroMechanical Systems (MEMS) sensors, including Inertial Measurement Units (IMU) and pressure sensors~\cite{fischer2021windnode}. Most commercial MEMS sensors are miniaturized System on Chip (SoC) devices featuring onboard analog-to-digital conversion~\cite{chen2023lopdm}, digital and analog programmable filters, and, in some cases, a machine learning core~\cite{zhang2021fault}. These systems enable on-device measurements and data processing, leveraging low-power processors like microcontrollers (MCU), while on-device feature extraction reduces network load for low-power, long-term installations~\cite{polonelli2022aerosense}. 

Existing wireless sensors monitoring large structures, such as wind turbines, typically employ a limited number of devices~\cite{fischer2021windnode,polonelli2022aerosense}. Recent works, namely the Aerosense system~\cite{barber2022development}, have proven the feasibility of installing \qty{4}{\milli\meter} thick, battery-supplied, and self-sustaining wireless sensor nodes directly on the surface of wind turbine blades, enabling long-term operative and structural monitoring analysis from an aerodynamic perspective. However, due to the substantial size of wind turbine blades (typically \qty{1}{\meter} to \qty{6}{\meter} chord length along a span of \qty{70}{\meter} for a \qty{5}{\mega\watt} wind turbine), deploying a large array of low-power, relatively high-frequency MEMS sensors on a large airfoil remains challenging. Ideally, a trade-off between measurement accuracy and power consumption should be found for more robust and long-term monitoring, with the optimal sensor selection and spatial distribution. In addition, the aerodynamic impact of any monitoring system installed on the surface of a wind turbine blade should be minimal, to avoid the system itself adversely affecting the power output of the wind turbine.

This paper is an extension of the conference paper~\cite{mistery_I2MTC} and focuses on comparing the performance of a non-intrusive system based on an array of 10 commercial MEMS pressure sensors to ground truth measurements from a high-accuracy pressure scanner in a wind tunnel, using a setup that emulates a wind turbine blade at 1:1 scale. The effect of the system on the blade aerodynamics is also evaluated by comparing experiments with and without it mounted on the blade. 
The experimental results demonstrate that the surface low-power MEMS sensor array system achieves comparable measurement accuracy to the reference pressure scanner setup,
The  measurement system features a total thickness of \qty{2.4}{\milli\metre}, comparable to the Aerosense system~\cite{barber2022development}, and a power consumption of \qty{1.6}{\milli\watt}. 

The results reported in this work confirm that MEMS pressure sensors are capable of measuring aerodynamic behavior at high Reynolds numbers on the surface of an airfoil in real-time without significantly affecting the airfoil aerodynamics. Therefore, we demonstrate that an ultra-thin (between 2 and \qty{3}{\milli\meter}) surface-mounted MEMS sensor array is suitable for structural health and condition monitoring on wind turbines without adversely affecting generation performance. 

To summarize, the main scientific contributions of this paper are: 
\begin{enumerate*}[label=(\roman*),,font=\itshape]
\item a quantitative comparison between a non-invasive low-power system, consisting of an array of compact low-cost MEMS pressure sensors on the surface of the blade, with invasive high-precision pressure scanners in a wind tunnel with high Reynolds number ($ Re = 3.5 \times 10^6 $);
\item investigation into the impact a \qty{2.4}{\milli\meter} thick system on the aerodynamics of an airfoil with a \qty{1.25}{\meter} chord and a span of \qty{5}{\meter} by comparing experiments with and without the MEMS pressure system;
\item discussion of the benefits and limitations of a plug-and-play MEMS pressure sensor system;
\end{enumerate*}

The main findings of this paper are that the performance of the MEMS sensors is comparable to that of high-accuracy lab-based pressure scanners, with a difference of \qty{2.3}{\%}. The total sensing power of the MEMS sensors is \qty{1.6}{\milli\watt}, allowing them to be used in the context of wind turbine blade monitoring. Additionally, the MEMS-based monitoring system does not significantly impact the aerodynamic performance of the wind turbine blades, with a maximum shift in the separation point of \ang{1}, making it suitable for integration without compromising efficiency.


\section{Related Works}
\label{sec:related_work}

MEMS pressure sensors have long shown promise as a lower-cost and lower-complexity aerodynamic monitoring solution, particularly in the aviation field~\cite{alsalem2023sensitivity}. In~\cite{raab_dynamic_2021}, dynamic flight tests were performed on a glider using a system of 64 MEMS pressure sensors fitted to a 'wing glove' placed around the wing chord-wise while increasing its cross-section slightly. The MEMS sensor data was compared to strain gauges fitted to the same glider and found to match qualitatively under various aerodynamic conditions, such as flow separation during stall maneuvers. The pressure sensor readings were also compared directly to an XFOIL simulation and found to match within \qty{50}{\pascal} after a linear regression scaling factor was applied. Limitations included the relatively bulky wing glove configuration and an uneven sensor placement that lacked sensors between the leading and trailing edges. A similar investigation has also been performed for fiber-optical pressure sensors~\cite{kienitz_static_2021}. The fiber-optic sensors used have a minimum thickness of \qty{1.6}{\milli\meter} and were compared with a Kulite LQ-080-258G thin line pressure transducer as a reference. The observed difference between the sensors was generally below \qty{5}{\hecto\pascal}. In the automotive sector, a sensing system consisting of a strip of 12 MEMS pressure sensors connected to an MCU has been investigated~\cite{zhang_automotive_2023}. Two experiments were performed: one in which the system was evaluated on a cylinder in a wind tunnel, and another in which the system was fitted to the inside of an S-duct, using a pressure scanner as a ground truth in both cases. The results showed that the MEMS absolute pressure sensors were comparable to pressure scanners in both scenarios, although notably, no investigation of the behavior of the system on an airfoil was performed. Moreover, the MEMS sensors used are not waterproof, making real-world deployment of this system impractical. 

Further sensing approaches using other kinds of MEMS sensors have also been investigated in the context of aircraft aerodynamics~\cite{alsalem2023sensitivity} and Structural Health Monitoring (SHM)~\cite{zonzini2020vibration}. For example, the authors of~\cite{ahlefeldt_road_2021} used 45 MEMS microphones to measure pressure fluctuations and compare their performance to traditional condenser microphones. Initially, investigations were performed in a wind tunnel at low speeds, before tests at transonic speeds in the wind tunnel and flight tests were performed. Large-scale MEMS microphone arrays have also been investigated in the context of aeroacoustics, such as in~\cite{ernst_enhancing_2024}. Modular panels containing 800 microphones each were mounted on the walls of a wind tunnel, resulting in an array of 7200 microphones \qty{6}{\meter} by \qty{3}{\meter} in size. The system was used to investigate noise emission sources and directivity. MEMS-based thermal wind speed sensors have also been investigated~\cite{wang2022influence} and found to measure with an accuracy of $\pm 7\%$ up to \qty{30}{\meter/\second} while detecting wind direction within $\pm$\qty{5}{\degree}. Another approach for detecting wind speed is using a compact spherical array of MEMS differential pressure sensors fitted to a drone and used for control purposes ~\cite{haneda_compact_2022}. Importantly, these investigations do not consider the high Reynolds numbers that occur during the operation of wind turbines, which typically range between $3$ and $15 \times 10^6$ for large installations~\cite{jung2022local}. High Reynolds numbers indicate turbulence which prevents a similarity scale approach from being used, since a scaled model would require wind speeds higher than the speed of sound to show similar effects, meaning  1:1 scale models are needed~\cite{zhu2021turbulence}. An investigation of MEMS absolute pressure sensor arrays under these circumstances has yet to be performed in the literature. 

In the context of wind turbine monitoring, the state of the art is given by the Aerosense system~\cite{barber2022development}, a self-sustaining monitoring system based on multi MEMS sensors and SoC MCU. The Aerosense system features pressure and acoustic MEMS sensor arrays and an IMU coupled with edge computing and photovoltaic energy harvesting to achieve in-field precise and sustainable data collection~\cite{polonelli2022aerosense}. Its thickness is in the \qty{4}{\milli\meter} range, a value similar to the setup used in this paper.
In particular, it is suitable for installation on a wide range of wind turbine blades hosting a modular sensor unit with up to 40 barometers and 10 microphones, in addition to 5 differential pressure sensors~\cite{polonelli2022aerosense}. 
%
Although the Aerosense system has been proven to run reliably on operating wind turbines~\cite{polonelli2022aerosense}, so far, works published in the literature mainly focus on the sensor design in terms of electronic components and installation~\cite{polonelli2022aerosense,barber2022development,polonelli2022calibration}. Moreover, the ultra-low power sensors selected for the Aerosense prototype have shown limitations regarding the sensor accuracy~\cite{polonelli2022calibration}, requiring long and tedious calibration procedures to reach the required absolute accuracy of \qty{10}{\pascal}~\cite{polonelli2022calibration}. Moreover, an investigation into the impact of Aerosense on the blade performance and a measurement accuracy assessment have never been performed~\cite{polonelli2022aerosense,barber2022development,polonelli2022calibration}.  

This paper extends the I2MTC conference paper~\cite{mistery_I2MTC} where we demonstrated the feasibility of inferring key aerodynamics parameters such as AoA from MEMS sensors placed on the surface of an operating wind turbine blade. As an extension, this paper presents a quantitative and punctual assessment of these low-power and cost-effective systems in terms of measurement accuracy and aerodynamic impact. This assessment is fundamental to compare such MEMS pressure sensors to industrial-standard ground truths demonstrating their accuracy in the context of wind turbine blade monitoring. 
Such an evaluation has yet to be performed in the literature, particularly at realistic Reynolds numbers for wind turbines of above $3 \times 10^6$, and this therefore forms one of the main contributions of this work. While the results described focus mainly on the use case of wind turbines, they have much broader implications for any aerodynamic applications where low-cost unobtrusive pressure data can be of interest, including aviation, automotive racing, and others. 

\section{Experimental Setup} 
\label{sec:setup}
\begin{figure*}
    \centering
    \includegraphics[width=.9\linewidth]{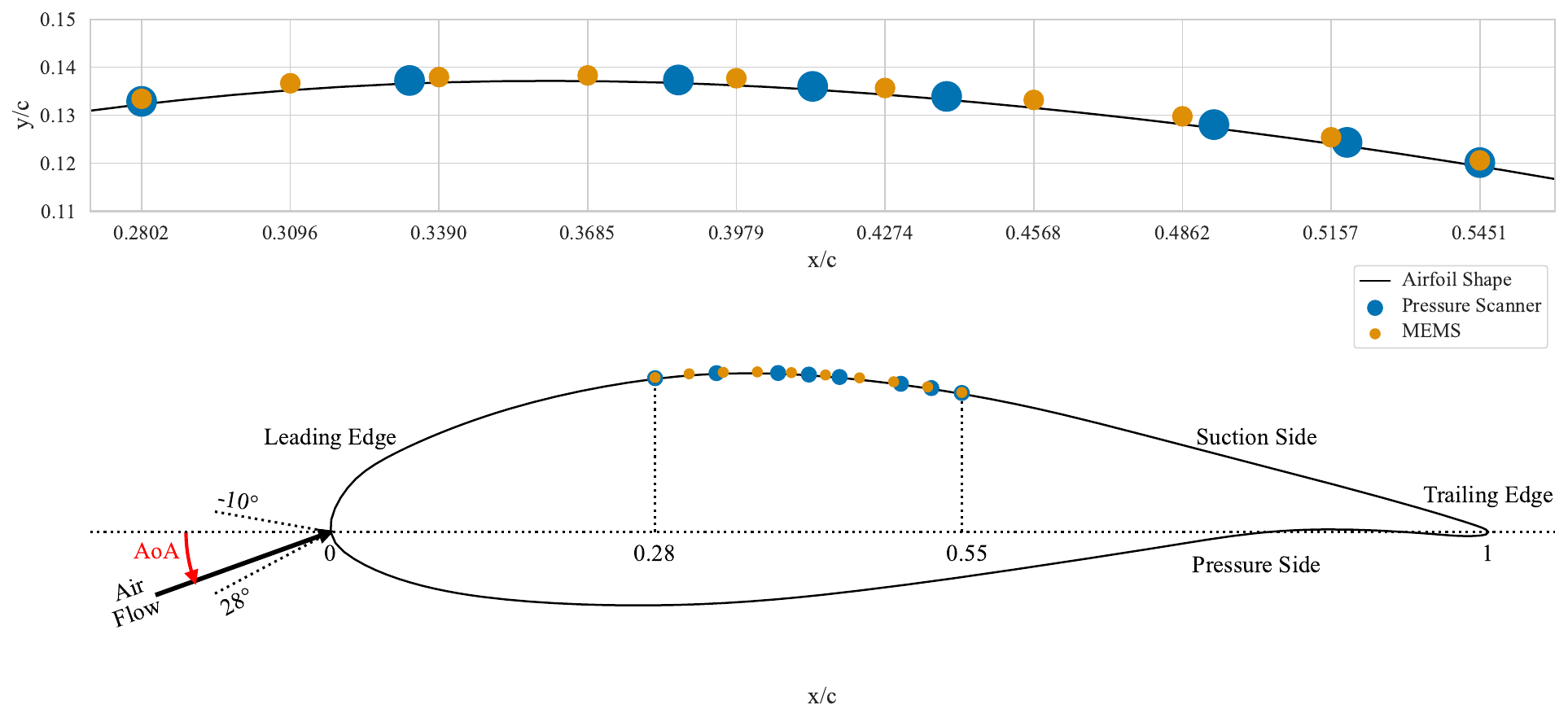}
    \caption{MEMS sensor and pressure tap distribution on the blade profile. This includes 10 evenly distributed absolute pressure MEMS sensors and 8 pressure taps.} 
    \label{fig:sensor_position}
\end{figure*}
\begin{figure}[t]
\centering
\centerline{\includegraphics[width=0.99\linewidth]{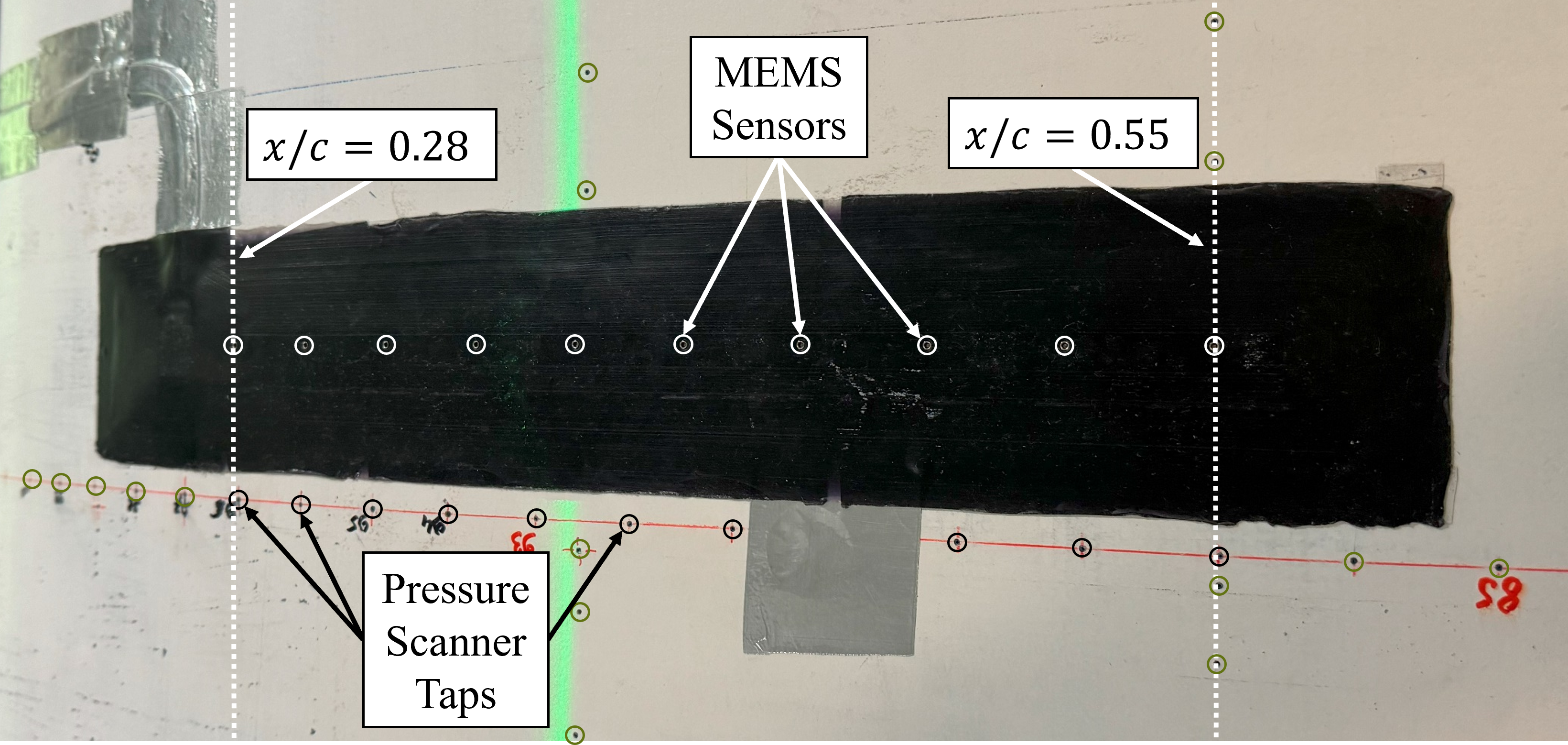}}
\caption{Top-down view of the encapsulated absolute pressure MEMS sensor array and pressure scanner taps on the blade surface. The encapsulation is the black rubber around the MEMS sensors.}
\label{fig:blade_photo}
\end{figure}

The experiments described in this paper were conducted in the Jules Verne wind tunnel owned by the Scientific and Technical Center for Building (CSTB) and located in Nantes, France.
The test section, measuring \qty{6}{\meter} in width and \qty{5}{\meter} in height, accommodates the airfoil with a \qty{1.25}{\meter} chord and a span of \qty{5}{\meter}. The maximum height is \qty{25}{\centi\meter}  (20\% of the chord). It occurs at 33\% of the chord length from the leading edge. The shape of the airfoil corresponds to a 1:1 scale of the section at 80\% of the span of an operating \qty{2}{\mega\watt} wind turbine blade~\cite{Neunaberetal2022}. The wind speed can reach up to \qty{60}{\meter\per\second}, reproducing the field conditions at a 1:1 scale. The AoA is varied using two jacks located at the extremities of the blade to be able to sustain \qty{2}{\tonne} of generated aerodynamic loading and maintain an angular precision below \ang{1}. The blade's structure has different hatches to install the acquisition system and wires inside the blade. The setup has a Reynolds number of $3.5 \times 10^6$ and a Mach number of 0.12 at \qty{40}{\meter/\second}.
On the blade surface, to measure the pressure distribution, a total of 350 high-quality pressure taps are distributed on the blade. This work focuses on a subset of these, placed in the vicinity of the MEMS sensor array, as seen on \Cref{fig:blade_photo} and \Cref{fig:sensor_position}. The MEMS acquisition system is described in more detail in \Cref{sec:sensors}.

A total of 18 tests were conducted, each consisting of two-minute acquisitions at a constant wind speed of \qty{40}{\meter\per\second} with the blade setup in \Cref{fig:sensor_position} and \Cref{fig:blade_photo}. Each acquisition has a varying AoA, from \ang{-10} to \ang{+28}, as described in~\cite{mistery_I2MTC}. In addition, between each AoA, short \qty{10}{\second} acquisitions are performed under stationary flow conditions. These experiments were also repeated without the MEMS sensor array present, allowing an evaluation of the aerodynamic effect of the monitoring system.  
\section{MEMS Sensing System Design and Ground Truth}
\label{sec:sensors}
This section describes the design of the MEMS absolute pressure sensor array system and its placement on the airfoil, as well as the configuration of the pressure scanner. Key sensor characteristics are described in \Cref{table:barometers}.
\subsection{MEMS Absolute Pressure Sensor Array}
The MEMS absolute pressure sensor array is designed to be placed on a key aerodynamic region of the blade, i.e., where the pressure varies the most when the AoA changes. Two pressure taps were chosen, at 28\% and 55\% as a reference, and ten MEMS sensors were placed equidistantly between them. This results in a spacing of \qty{37}{\milli\meter}.
The MEMS sensor selected is the LPS28DFW from STMicroelectronics; it is an evolution of the sensor used on Aerosense (LPS27HHW) \cite{polonelli2022aerosense}, providing $1.7\times$ better relative accuracy and a power consumption reduction of 28\%. The improved relative accuracy reduces the need for the tedious calibration process previously required~\cite{polonelli2022calibration}. The sensor is also waterproof, enabling real-world deployment in the future.

To accommodate the sensor array, a Printed Circuit Board (PCB) was designed and deployed on the blade surface with a maximum component thickness of \qty{1.95}{\milli\meter}. The substrate is \qty{0.2}{\milli\meter} Polyimide coupled with a \qty{0.2}{\milli\meter} stiffener, featuring complete flexibility to adhere to the blade surface. The PCB is based on the STM32L452, and supports an array of ten LPS28DFW. The board is encapsulated using custom 3D-printed flexible rubber, see \Cref{fig:blade_photo}. This is done to smoothen the surface to limit the impact of sharp edges of the PCB and electronic components on the airflow, which could skew the measurements. The maximum thickness of the MEMS pressure measurement system is \qty{2.4}{\milli\meter}, i.e., \qty{0.96}{\%} of the blade height. The power consumption for an array of ten LPS28DFW sampling at \qty{100}{\hertz} is \qty{1.6}{\milli\watt}.

\subsection{Pressure Scanners}
\label{subsec:pressure_scanner}
Small holes (pressure taps) of \qty{0.8}{\milli\meter} in diameter are drilled through the 1:1 scale wind turbine blade to measure the surface pressure generated by the flow passing over the airfoil, which generates the lift force.
These pressure taps are connected to multiplexed ESP pressure scanners from TE Connectivity, using \qty{1.5}{\meter} long vinyl tubes. They measure differences of pressure between the airfoil ($P_i$) and the free airstream in the wind tunnel ($P_\infty$), as defined in \Cref{eq:deltap}.
\begin{equation}
\label{eq:deltap}
\Delta P_{scanner} = P_{i} - P_{\infty}
\end{equation}
A pressure scanner consists of an array of silicon piezoresistive pressure sensors, designed to maintain long-term stability and provide accurate measurements within \(\pm 0.03\%\) of static errors over the full-scale pressure range. The sensors operate effectively from \qty{0}{\kilo\pascal} to \qty{7}{\kilo\pascal} near the leading edge suction peak and from \qty{0}{\kilo\pascal} to \qty{2.5}{\kilo\pascal} in other regions.

The transfer function of the whole system, i.e., tubes and sensor cavity, has been characterized offline at a sampling frequency of \qty{1024}{\hertz} and has been taken into account in the processing of the pressure data. During the measurements, the sampling frequency was \qty{512}{\hertz}. The signal acquisition was performed using two National Instrument acquisition boards synchronized using a dedicated \qty{1}{\hertz} signal. 
\begin{table*}[t]
\centering
\caption{Sensors selected for the experiments reported in \Cref{sec:results}.}
\label{table:barometers}
\setlength{\tabcolsep}{9pt}
\begin{tabular}{l l c c c c}
\hline\hline
\textbf{Type} & \textbf{Model} & \textbf{Quantity} & \textbf{Sampling Rate [\si{\hertz}]} & \textbf{R/A Accuracy$^\diamond$[\si{\pascal}}] & \textbf{Height$^\circledast$ [\si{\milli\meter}]} \\
\hline
MEMS absolute pressure sensors & ST LPS28DFW & 10 & 100 & 1.5 / 50 & 1.95 \\ 
Pressure scanners & TE ESP & 8 & 512 & - / 7.5 & 0 \\ 
\hline\hline

\multicolumn{6}{l}{ $^\diamond$ Relative/Absolute accuracy, $^\dagger$ Cumulative generated bitrate, $^\circledast$ Extra height added on top of the blade surface} \\
\end{tabular}
\end{table*}
\section{Data Processing and Normalization}
\label{sec:data_proc_and_norm}
Before comparing the acquired measurements, several data processing steps are required. First,  a differential outlier filter is applied to the absolute pressure data from the ten LPS28DFW barometers to eliminate outliers. In \Cref{eq:deltamems} the atmospheric pressure  ($P_{atm}$) is removed from the MEMS absolute pressure sensor acquisitions by subtracting the mean of ten-second measurements at stationary flow conditions. 
\begin{equation}
\label{eq:deltamems}
\Delta P_{MEMS} = P_{i} - P_{atm}
\end{equation}
As described in \Cref{subsec:pressure_scanner}, the measurements from the pressure scanner are given by \Cref{eq:deltap}, where $P_{i}$ becomes the pressure on the blade surface and $P_{\infty}$ is the airstream pressure.
Following the Bernoulli formula along a streamline, \Cref{eq:deltap} formulates as \Cref{eq:atm}.
\begin{equation}
\label{eq:atm}
P_{atm} + 0 = P_\infty+q_\infty
\end{equation}
Therefore the pressure measured by the MEMS sensors can also be written as \Cref{eq:mems}, where $q_{\infty} = 0.5 \rho U^2_\infty$ is the so-called dynamic pressure, $\rho$ is the air density, and $U_\infty$ the upstream wind velocity.
\begin{equation}
\label{eq:mems}
\Delta P_{MEMS} = P_{i} - \left( P_{\infty} + q_{\infty}\right)
\end{equation}
Although the pressure scanner and the MEMS sensors measure slightly different quantities, it is therefore possible to compare them  
by adding the dynamic pressure $q_{\infty}$ as described in \Cref{eq:scannfinal}. A calibration coefficient $\alpha$ is introduced. It is chosen to align the data for $AoA = 24$. In this case, $\alpha = 0.9$
\begin{equation}
\label{eq:scannfinal}
\Delta P_{scanner,\text{ from MEMS}} := \Delta P_{MEMS}+q_{\infty} \times \alpha
\end{equation}
In the following results and figures, the MEMS absolute pressure sensor array results are shown with the pressure as defined here throughout.
\section{Experimental Results}
\label{sec:results}
This section presents the results in three stages. First, it details the data obtained from the MEMS sensor system. Second, it provides a comparison with the pressure scanner data. Finally, it analyzes the impact of the MEMS system on the aerodynamics of the blade by comparing the acquisition process with and without the MEMS system attached to the blade surface.

\subsection{MEMS sensors system data analysis}
\begin{figure}[t]
\centering
\centerline{\includegraphics[width=0.93\linewidth]{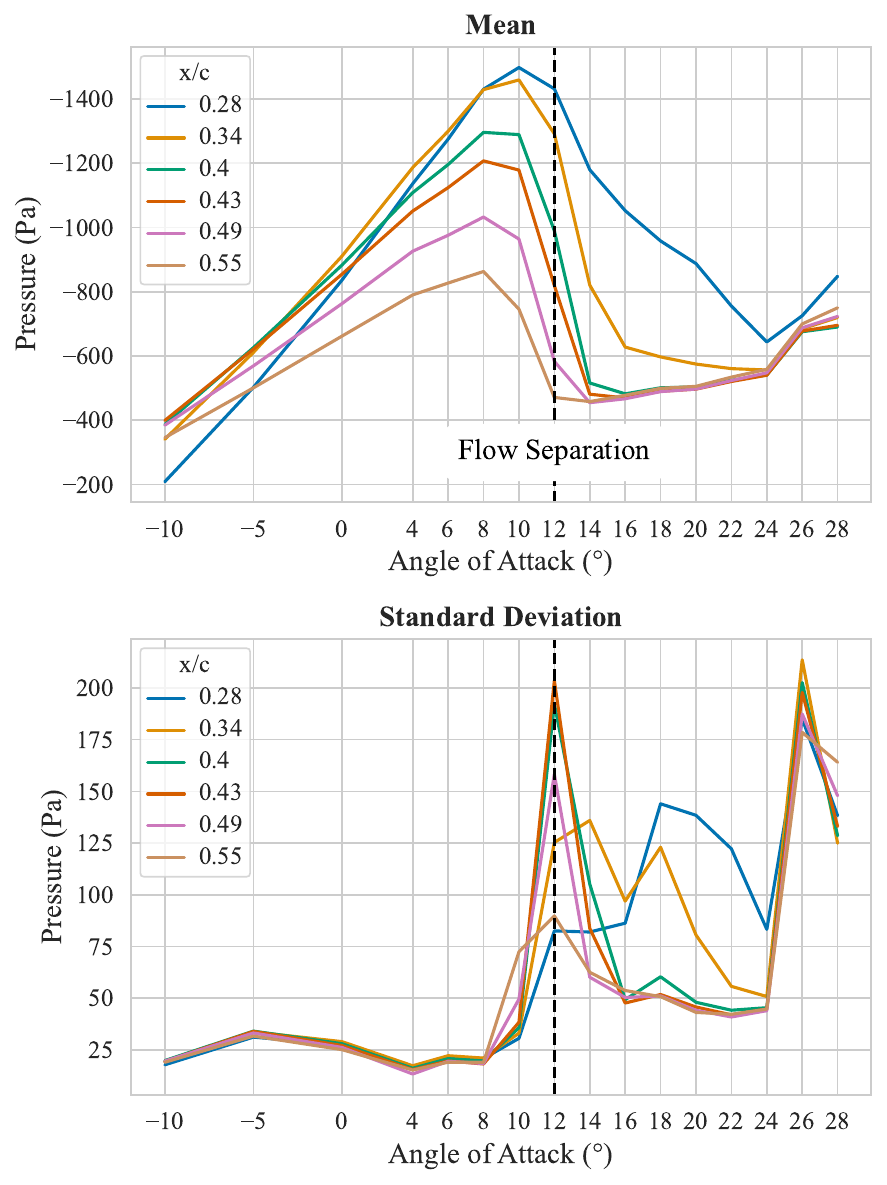}}
\caption{Inverted mean and standard deviation over time of the absolute pressure sensors (in Pa) depending on the blade AoA.}
\label{fig:abs_ps}
\end{figure}

The mean and standard deviation over time of the MEMS pressure sensor array for different angles of attack is shown in \Cref{fig:abs_ps}. 
\begin{figure}[t]
    \centering
    \includegraphics[width=0.93\linewidth]{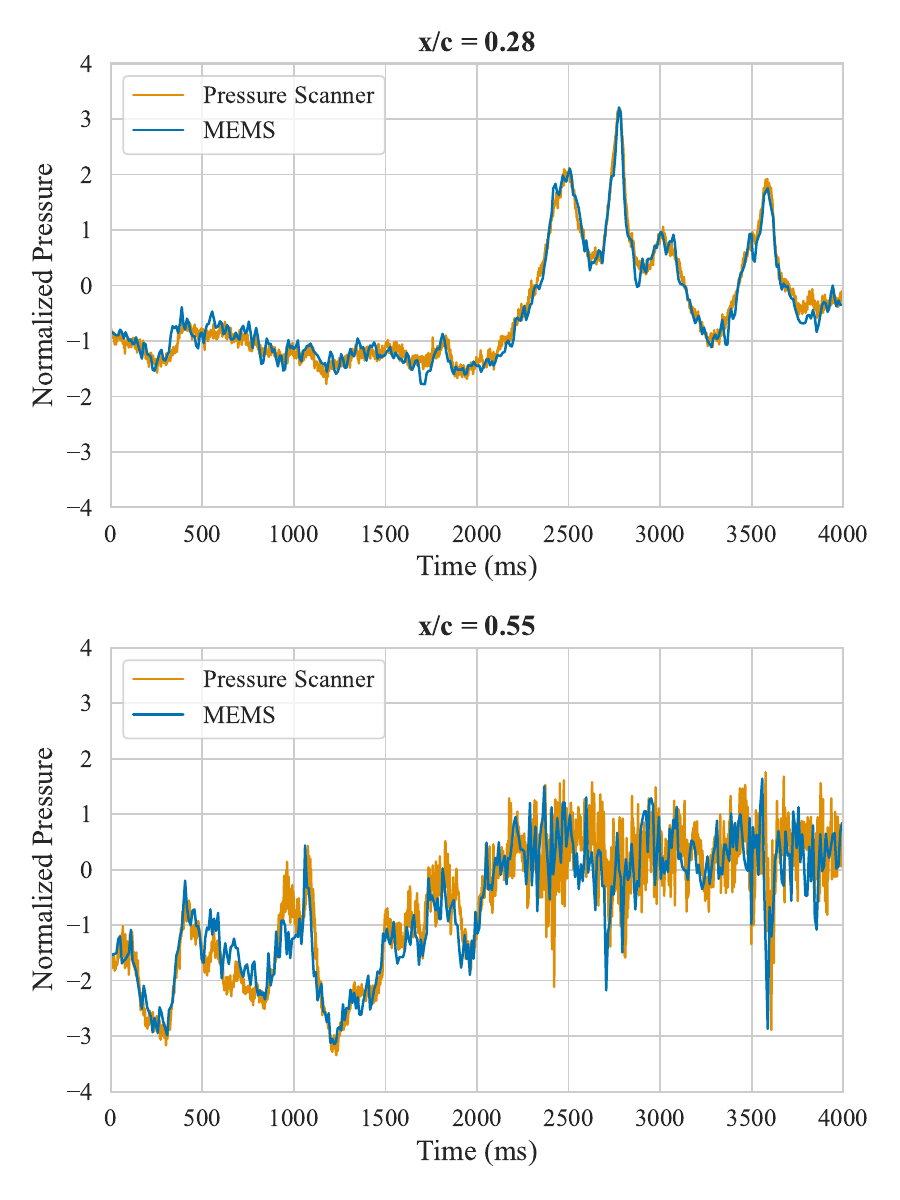}
    \caption{Normalized raw sensor data for pressure scanners and MEMS sensors for \( x/c = 0.28 \) (top) and \( x/c = 0.55 \) (bottom).}
    \label{fig:ext_sync_data}
\end{figure}
\begin{figure}[t]
    \centering
    \includegraphics[width=0.93\linewidth]{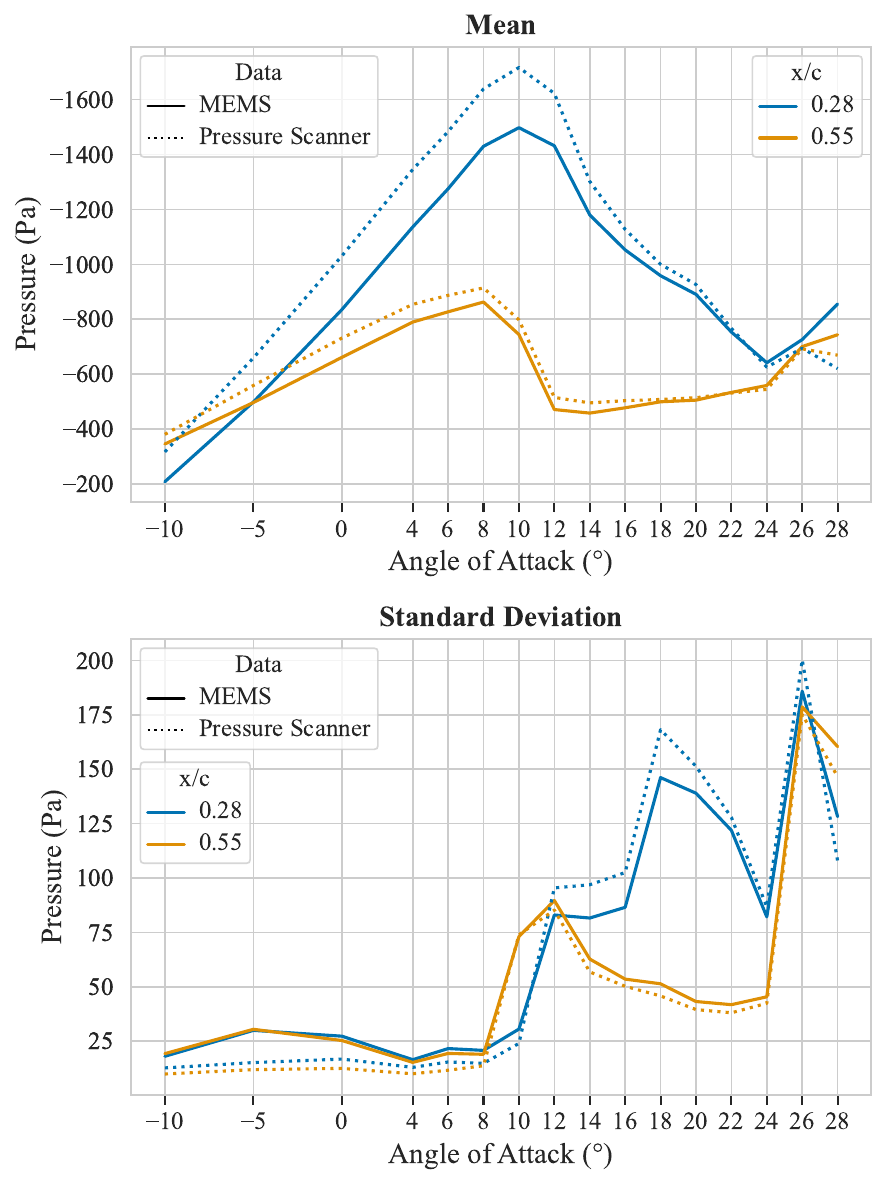}
    \caption{Inverted mean pressure readings (top) and standard deviation (bottom) over time for different angles of attack for the two furthest apart sensor pairs. Dotted lines represent pressure scanner data and plain lines represent MEMS sensors.}
    \label{fig:ext_MeanStd}
\end{figure}
\begin{figure}[t]
    \centering
    \includegraphics[width=0.93\linewidth]{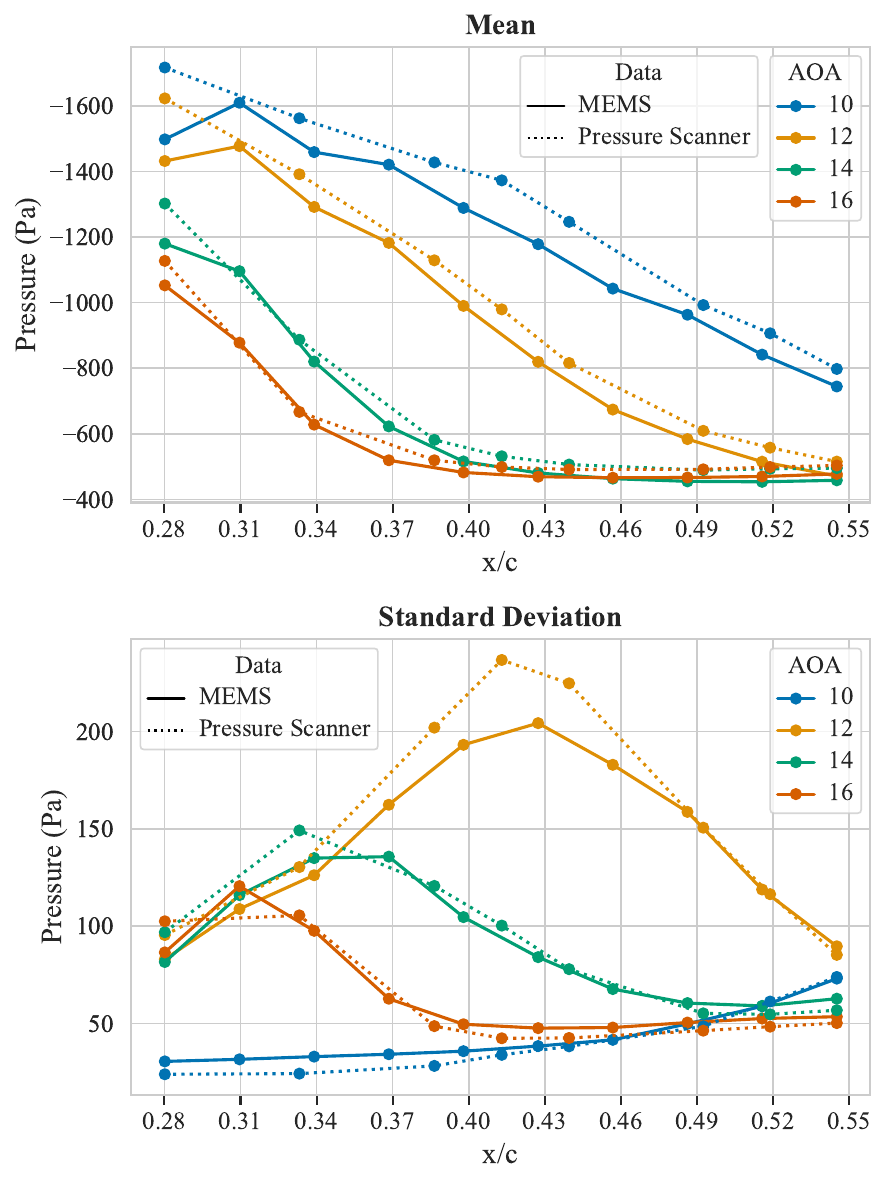}
    \caption{Inverted mean pressure readings (top) and standard deviation (bottom) over time along the chord length for measured angles of attack between \qty{+10}{\degree} and \qty{+16}{\degree}. Dotted lines represent pressure scanner data and plain lines represent MEMS sensors.}
    \label{fig:ext_MeanStd_xc}
\end{figure}
The aerodynamic force comes mainly from a generated suction on the upper side of the airfoil (hereafter called the suction side), which means the sensors measure negative pressure relative to the atmosphere. An inversion of the y-axis on the mean measurements allows for a more direct interpretation of the generated suction and therefore of the lift forces acting on the airfoil. That way, the maximum lift generated is represented when the value reaches its maximum peak (i.e. the minimum pressure). When the AoA is increased, the suction increases (i.e., the pressure decreases) linearly with the AoA, an effect clearly visible in \Cref{fig:abs_ps} between \ang{-10} and \ang{8}. This result is confirmed by previous experiments in the same conditions and on the same airfoil~\cite{braud2023study}, where a maximum suction is found at \ang{10} before flow separation. Above \ang{10}, the flow on the suction side starts to separate from the trailing edge which reduces the overall suction generated on the airfoil. The flow starts to separate at the trailing edge, and as the AoA is increased, the limit (called the separation point) between the attached flow region near the leading edge and the separated flow region near the trailing edge, moves upstream. This effect is visible from the data collected by the array of ten absolute pressure sensors, depicted in \Cref{fig:abs_ps}, where the suction drops down to a local plateau at about \qty{-500}{\pascal}, starting with the pressure sensors closer to the trailing edge. Moreover, the effect of flow separation intrinsically generates increased local fluctuations, which are visible with the rise in standard deviation around \ang{12} of \Cref{fig:abs_ps}. Notably, the suction decreases more slowly for the pressure sensors closer to the leading edge, e.g., \( x/c = 0.28 \) and \( x/c = 0.34 \), than for the ones to the trailing edge side, e.g.,\( x/c = 0.49 \) and \( x/c = 0.55 \), between \ang{10} and \ang{26}. For AoA larger than \ang{26}, the flow becomes fully separated from the airfoil and reaches another state, which can be seen by another peak of the standard deviation. 

\subsection{Comparison of MEMS sensors and high-accuracy pressure sensors}

\Cref{fig:ext_sync_data} compares synchronized and normalized raw signals for the pressure scanners and the MEMS sensors at \( x/c = 0.28 \) and \( x/c = 0.55 \) for \( AoA = \qty{12}{\degree} \). These are the sensors closest to the leading edge and closest to the trailing edge respectively. The AoA was chosen because the flow is separated at the trailing edge and at the pressure tap \( x/c = 0.55 \), but attached at the leading edge and at the pressure tap \( x/c = 0.28 \). This plot provides information about the dynamic response of the sensors. The MEMS sensors follow the pressure scanner accurately at lower frequencies, i.e., when the flow is attached. However, when higher frequencies arise, i.e., when the flow is separated, the lower sampling frequency of the MEMS sensors can be observed. However, the filtered signal from the pressure scanners is similar to the MEMS time series.

\Cref{fig:ext_MeanStd} presents a comparison of the sensors for time-averaged behavior analysis, it shows the mean and standard deviation for the same pair of MEMS sensors and pressure taps as \Cref{fig:ext_sync_data}, i.e., \( x/c = 0.28 \) and \( x/c = 0.55 \) across all the AoA averaged over time.
For both sensor locations at \( x/c = 0.28 \) and \( x/c = 0.55 \) the mean follows the expected linear trend for low angles of attack \ang{-10} to \ang{8}, however, a slight offset is present. While the sensors close to the trailing edge, i.e., \( x/c = 0.55 \), show an offset between 35 to \qty{70}{\pascal}, the sensors close to the leading edge, i.e., \( x/c = 0.28 \), show an increasing offset from 110 to \qty{210}{\pascal}. The offset in the standard deviation for low AoA remains below \qty{19}{Pa}. At the AoA of \ang{10} the sensors closer to the trailing edge, at \( x/c = 0.55 \), show the expected decrease in the suction alongside a sudden increase in standard deviation due to the flow separation, as previously discussed. The sensors at \( x/c = 0.28 \) are not yet affected at \ang{10}, showing the same expected behavior at \ang{12}, due to the flow separation point moving towards the leading edge with increasing AoA. The largest offset between the pressure scanner and the MEMS pressure sensors at \( x/c = 0.55 \) with \qty{220}{\pascal} occurs at the same AoA as the suction peak at \ang{10}. For larger AoA, i.e., more than \ang{12}, the absolute suction offset between the MEMS sensors and the pressure scanner for both sensor pairs steadily decreases and only increases again after the flow fully separates at \ang{26}. Comparing the standard deviation for larger AoA, the offset stays low around \qty{5}{\pascal} for \( x/c = 0.55 \) while fluctuating between \qty{5}{\pascal} and \qty{22}{\pascal} for \( x/c = 0.28 \). For the fully separated flow after \ang{26} the offset in the standard deviation increases again. Summarizing the presented plot on the static behavior analysis of the MEMS and pressure scanner sensors, the MEMS sensors follow the behavioral trends of the pressure scanners overall, however a slight offset is present. The offset affects the sensors near the leading edge more strongly than the sensors near the trailing edge and demonstrates changing behavior for different AoA-regions: low AoA from \ang{-10} to \ang{8} show a fluctuating or increasing offset, high AoA \ang{12} to \ang{24} show a decreasing offset, very high AoA above \ang{26} show an increasing offset again.

\Cref{fig:ext_MeanStd_xc} presents the mean and standard deviation over the chord length for the angles of attack between \qty{+10}{\degree} and \qty{+16}{\degree} for the MEMS sensors and pressure scanner. The markers show the exact positions of the MEMS sensors and the pressure taps along the chord. It can be seen that for an \(AoA=\qty{10}{\degree}\) and \(AoA=\qty{12}{\degree} \) the pressure gradient is constant for each angle and follows the same slope for both the MEMS array and the pressure scanner.
For all the AoA, there is a plateau around \qty{-500}{\pascal}. However, it starts closer to the leading edge as the AoA increases. This is produced by the region of separated flow.

Comparing the presented mean and standard deviation of the MEMS and pressure scanner sensors, it is observed that an increase of the AoA within the presented selection leads to a reduction of the offset, as well as that differences appear worse near the leading edge than near the trailing edge, just as deduced from \Cref{fig:ext_MeanStd}. However, it is more complex to estimate a clear difference between the measurements along the chord due to the different positions of the MEMS sensors and pressure taps. For example, around the position \( x/c = 0.31 \) some of the biggest differences seem to occur between the measurements. However, this blade region is covered by a limited number of pressure taps, only allowing a limited comparison of the linear spatial approximations. On the bottom plot, the standard deviation graph indicates that both systems equally capture the separation point moving from the trailing edge towards the leading edge for higher AoA during the flow separation, shown by a gradual shift of a bumpy peak in the distribution. Already for \qty{+10}{\degree} an increase of the standard deviation towards the trailing edge can be observed, hinting at a separation point slightly outside of the chord scope. The intermittent separation point, defined as the location of the peak fluctuation maximum, can be estimated at roughly the same locations for the MEMS and pressure scanner sensors, despite the pressure scanners featuring more extreme peaks. Due to the sparse and different distribution of the sensors, a more detailed comparison of the intermittent separation point locations is restricted. However, similar trends and relations are evident.

As explained in \Cref{sec:sensors}, not all the MEMS sensors have a pressure tap in their vicinity. However, a clear matching trend between the two types of sensors can be observed.
The behavior is similar to the one observed in~\cite{braud2023study}. These plots show that it is possible to rely on low-power compact MEMS sensor data for steady pressure analysis.

\subsection{Impact of the MEMS measurement system on the blade aerodynamics}
\begin{figure}[t]
    \centering
    \includegraphics[width=0.93\linewidth]{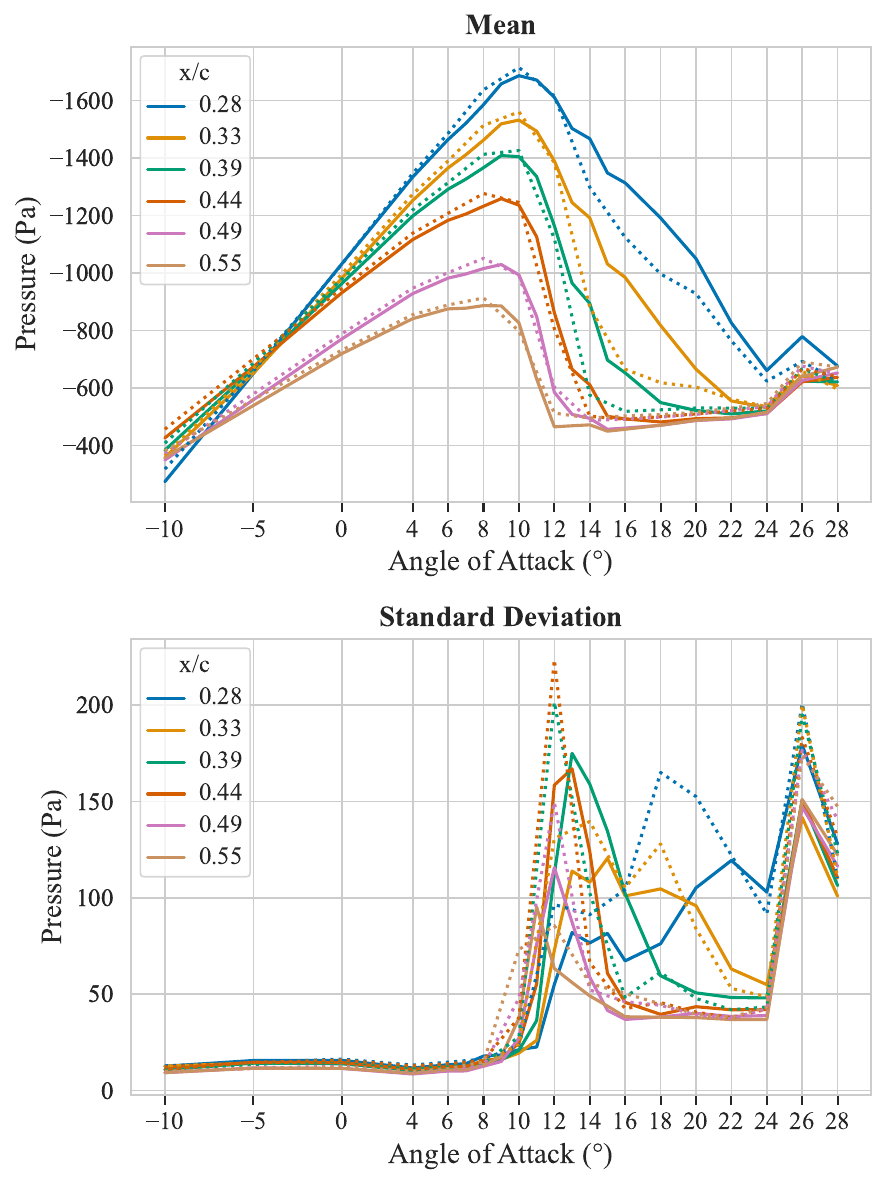}
    \caption{Inverted mean (top) and standard deviation (bottom) of pressure scanners between \( x/c = 0.28 \) and \( x/c = 0.55 \) for AoA between \qty{-10}{\degree} and \qty{+28}{\degree}. Full lines represent data gathered before mounting the MEMS sensor system and dotted lines represent the data when the MEMS system is added on the surface of the blade, next to the pressure taps as described in \Cref{fig:blade_photo}.}
    \label{fig:ext_J3vsJ4_MeanStd}
\end{figure}
To evaluate the relevance of adding sensor nodes on wind turbine blades, it is important to estimate the impact of such devices on the aerodynamics in their vicinity. This provides an idea of the changes in the generated lift caused by such a monitoring system installed on an airfoil. The encapsulation of the system, described in \Cref{sec:sensors},  slowly increases in thickness from its edge until it reaches the thickness of the MEMS pressure sensors, i.e., \qty{1.95}{\milli\meter}. The housing is \qty{48.5}{\centi\meter} long, \qty{8.5}{\centi\meter} wide and \qty{2.4}{\milli\meter} thick.

\Cref{fig:ext_J3vsJ4_MeanStd} describes a comparison between data acquisition from the pressure scanner before and after the additional installation of the MEMS sensor array on the blade. The pressure taps of the pressure scanner are located parallel to the encapsulation of the system, at \qty{8.5}{\mm} from the edge. It shows the mean and standard deviation for the pressure over different angles of attack for a selection of locations along the chord between \( x/c = 0.28 \) and \( x/c = 0.55 \). Looking at the upper plot, one can observe that the two acquisitions share very similar results before the start of flow separation at the trailing edge around (\(AoA=\qty{10}{\degree} \)). 
With the addition of the MEMS sensors on the blade, during the transition from a fully attached flow \((AoA<\qty{10}{\degree} \)) to a fully separated flow (\(AoA>\qty{26}{\degree}\)), the decreasing slope of the pressure seems more abrupt, while the AoA where the maximum of suction occurs stay identical.

Observing the standard deviation, for low angles of attack and an attached flow (\(AoA<\qty{8}{\degree}\)) the measurements match closely. However, then the acquisition including the addition of the MEMS sensor system appears to show earlier symptoms of flow separation, presenting earlier peaks than the acquisition without the added system. This offset goes from \qty{0}{\degree} for \( x/c = 0.49 \) to \qty{1}{\degree} for \( x/c = 0.44 \). Towards high angles of attack both standard deviation measurements peak again at the \(AoA=\qty{26}{\degree} \) where the flow becomes fully separated. The acquisition with the MEMS system on top exhibits higher peaks of fluctuations than the acquisition without them.
\section{Discussion}
\label{sec:discussion}
From the MEMS sensors alone, it can be seen that it may be possible to infer the AoA and the flow separation of an operating wind turbine with low-cost and low-power pressure sensors measuring directly from the blade surface. Moreover, \Cref{fig:abs_ps} suggests that there is a strong correlation among the sensors, indicating a potential redundancy in the measurements. Results are detailed in \Cref{table:results}.
\begin{table}[t]
\centering
\caption{Average error for mean and standard deviation over all angles of attack for \( x/c = 0.28 \) and \( x/c = 0.55 \) as described in \Cref{fig:ext_MeanStd}.}
\label{table:results}
\setlength{\tabcolsep}{6pt}
\begin{tabular}{l c c}
\hline\hline
\textbf{Sensor position} & \textbf{Mean error (\%)} & \textbf{Standard deviation error (\%)} \\
\hline
\textbf{\( x/c = 0.28 \)} & 7.5 & 5.5 \\
\textbf{\( x/c = 0.55 \)} & 2.3 & 3.3 \\
\hline\hline
\end{tabular}
\end{table}
%

The comparison between the MEMS sensor array and the pressure scanner shows a small, but varying offset between the two sensing approaches. \Cref{table:results} shows that for \( x/c = 0.55 \), the average error over the AoA is \qty{2.3}{\%} for mean and \qty{3.3}{\%} for standard deviation. However, the error is higher for the sensor at \( x/c = 0.28 \).
There are a variety of possible causes for this. Aside from the slight differences in sensor positions and potentially slightly lower accuracy of the MEMS sensors compared to the pressure scanner, aerodynamic causes are also not unlikely. On one side, the housing and the encapsulation of the MEMS sensor array may be affecting their readings or the aerodynamics of the blade. It is also possible that the assumptions made in \Cref{sec:data_proc_and_norm} to enable the comparison between the MEMS sensors and the pressure scanner may not entirely hold. In particular, small inaccuracies in the measured wind velocity and calculated air density may cause the offset observed. Alternatively, it is possible that the MEMS sensors are not impacting the aerodynamics of the blade itself, but rather affecting the readings recorded by the pressure scanner due to their proximity, while the blade aerodynamics in itself remains largely unchanged. 

The results of the comparison between the pressure scanner measurements with and without the presence of the MEMS sensors indicate that flow separation may be triggered slightly faster when the MEMS sensor array is installed than for a clean blade. However, when the flow starts to separate, the flow becomes more three-dimensional, and the flow is no longer entirely uniform along the span. Therefore the local extra thickness due to the encapsulation of the sensor may trigger 3D structures in the flow, which may or may not exist without the extra thickness. The influence of the sensors' housing may be reduced with a system that fully covers the blade and does not create a local extra thickness, therefore increasing its cross-section. However, this increase in cross-section is in itself undesirable. 

Overall, the results suggest that MEMS pressure sensor arrays are a promising approach to monitoring wind turbine blades, and provide comparable results to high-accuracy pressure scanners in a wind tunnel environment. 

\section{Conclusion}
\label{sec:conclusion}
Monitoring wind flow and operational parameters of wind turbines directly in the field is a fundamental prerequisite for optimizing sustainable electrical generation efficiency. The aerodynamics of the rotating blades is of particular interest; however, many technical and scientific challenges still exist due to the complexity of measurement system installations.
This paper compares a monitoring system featuring a non-invasive, low-cost, surface-mounted MEMS absolute pressure sensor array with a high-accuracy pressure scanner on a 1:1 scale wind turbine blade with Reynolds number $3.5 \times 10^6$ in a wind tunnel. The accuracy of the MEMS array is compared with the pressure scanner and the effects of the MEMS array on the aerodynamics of the turbine blade are investigated. 
The results show that the MEMS pressure sensor array produces comparable readings to the pressure scanner, with an error as low as \qty{2.3}{\%}. 
The MEMS sensor array causes flow separation to occur at slightly lower AoA, i.e., up to \ang{1} less than when only the pressure scanner is present. Otherwise, it does not appear to meaningfully affect the blade aerodynamics, especially in the region of interest for the application targeted, i.e., before flow separation. With an ultra-low power consumption of \qty{1.6}{\milli\watt}, it proves the relevance of MEMS sensors for wind turbine blade monitoring.  Future work will investigate the cause of the observed offset further, evaluate the MEMS array system during more unsteady airflows at high Reynolds numbers, and analyze the practical implications and possibilities of deploying a MEMS pressure sensor array for wind turbine monitoring. 

\bibliography{bib, Mistery_TIM_Extension_DM}{}
\bibliographystyle{IEEEtran}


\end{document}